\documentclass[amssymb,aps,psfig,epsf]{revtex4}

\usepackage{graphicx}
\begin{document}
\title{Odd Triplet Superconductivity in Superconductor/Ferromagnet System with a
Spiral Magnetic Structure. \\
}
\author{A.F. Volkov$^{1,2},$ A. Anishchanka$^{1}$ and K.B.Efetov$^{1,3}$}
\address{$^{(1)}$Theoretische Physik III,\\
Ruhr-Universit\"{a}t Bochum, D-44780 Bochum, Germany\\
$^{(2)}$Institute for Radioengineering and Electronics of Russian Academy of%
\\
Sciences,11-7 Mokhovaya str., Moscow 125009, Russia\\
$^{(3)}$L. D. Landau Institute for Theoretical Physics RAS, 119334 Moscow,\\
Russia}

\begin{abstract}
We analyze a superconductor-ferromagnet (S/F) system with a spiral magnetic
structure in the ferromagnet F for a weak and strong exchange field. The
long-range triplet component (LRTC) penetrating into the ferromagnet over a
long distance is calculated for both cases. In the dirty limit (or weak
ferromagnetism) we study the LRTC for conical ferromagnets. Its spatial
dependence undergoes a qualitative change as a function of the cone angle $%
\vartheta $. At small angles $\vartheta $ the LRTC decays in the ferromagnet
exponentially in a monotonic way. If the angle $\vartheta $ exceeds a
certain value, the exponential decay of the LRTC is accompanied by
oscillations with a period that depends on $\vartheta $. This oscillatory
behaviour leads to a similar dependence of the Josephson critical current in
SFS junctions on the thickness of the F layer. In the case of a strong
ferromagnet the LRTC decays over the length which is determined by the wave
vector of the magnetic spiral and by the exchange field.
\end{abstract}
\maketitle

\section{Introduction}

\bigskip

It is well known that superconductivity and ferromagnetism are antagonistic
phenomena (see, for example, the reviews \cite{BulR,GolubovR,BuzdinR}).
Exchange interaction in ferromagnets results in ordering electron spins in
one direction, whereas superconducting correlations in conventional
superconductors lead to the formation of Cooper pairs with opposite spins of
electrons. The antagonistic character of ordering in ferromagnets (F) and in
superconductors (S) is the reason for an essential difference between the
proximity effects in S/N and S/F structures (here N denotes a normal
nonmagnetic metal). In S/N structures the condensate penetrates the normal
metal N over a rather long distance which in the dirty limit ($\tau T<<1,$ $%
\tau $ is the elastic scattering time) is equal to

\begin{equation}
{\ }\xi _{N}=\sqrt{D/2\pi T}  \label{I1}
\end{equation}
where $D=vl/3$ is the diffusion coefficient and $l=v\tau $ is the mean free
path. On the other hand, the depth of the condensate penetration into a
ferromagnet in S/F system is much shorter

\begin{equation}
{\ }\xi _{F}=\sqrt{D/h}  \label{I2}
\end{equation}
if the exchange energy $h$ is larger than the temperature $T$ (usually $h>>T$%
). The formula (\ref{I2}) is valid for a short mean free path (${\ }h\tau
<<1 $). If the exchange field $h$ is strong enough ($h\tau >>1$), the
condensate penetrates the ferromagnet over a distance of the order of $l$
(if $\tau T<1$) and oscillates with the period $\sim v/h$ \cite
{BVEpr01,BuzdinBal}.

Note that a short length of the condensate penetration is related to the
fact that Cooper pairs in a conventional superconductor are formed by two
electrons with opposite spins. In the case of a homogeneous magnetization
the wave function $f(t-t^{\prime })$ of Cooper pairs penetrating into the
ferromagnet consists of two parts

\begin{eqnarray}
&&f_{3}(t-t^{\prime })\sim \langle \psi _{\uparrow }(t)\psi _{\downarrow
}(t^{\prime })-\psi _{\downarrow }(t^{\prime })\psi _{\uparrow }(t)\rangle
\label{I5} \\
&&f_{0}(t-t^{\prime })\sim \langle \psi _{\uparrow }(t)\psi _{\downarrow
}(t^{\prime })+\psi _{\downarrow }(t^{\prime })\psi _{\uparrow }(t)\rangle
\label{I6}
\end{eqnarray}

The first function describes the singlet component. It differs from zero
both in the whole superconducting region and in the ferromagnet over the
length $\xi _{F}$. The second function $f_{0}(t-t^{\prime })$ describes the
triplet component with zero projection of the magnetic moment of a Cooper
pair on the $z$-axis (the magnetization vector ${\bf M}$ is oriented along
the $z$-axis). It is not zero only in the vicinity of the S/F interface,
over a distance $\sim \xi _{S}=\sqrt{D/2\pi T_{c}}$ in the superconductor
and over the distance $\xi _{F}$ in the ferromagnet. This function is an odd
function of the difference $(t-t^{\prime })$ and therefore is equal to zero
at $t=t^{\prime }.$ This means that in the Matsubara representation $%
f_{0}(\omega )$ is an odd function of $\omega $, whereas the function $%
f_{3}(\omega )$ is an even function of $\omega $.

All the statements above concern only the case of a homogeneous
magnetization in the F region. The situation changes qualitatively if the
magnetization is not homogeneous, for example, if one has a spiral magnetic
structure in the ferromagnet. In this case not only the singlet and triplet
component with zero projection of magnetic moment, but also a triplet
component with a nonzero projection of magnetic moment arises in the system.
This type of the triplet component means that Cooper pairs appear in the
system that are described by the condensate function $%
f_{tr}(t-t^{\prime })\sim \langle \psi _{\uparrow }(t)\psi _{\uparrow
}(t^{\prime })\rangle $ or $f_{tr}(t-t^{\prime })\sim \langle \psi
_{\downarrow }(t)\psi _{\downarrow }(t^{\prime })\rangle .$ The triplet
pairing is well known in superfluid $He^{3}$ \cite{Legget,Wolfle} or in $%
Sr_{2}RuO_{3}$ \cite{Maeno} and is believed to be realized in compounds with
heavy fermions \cite{Mineev}. The condensate function $f_{tr}(\omega )$ in
these materials is an odd function of momentum ${\bf p}$ and an even
function of the Matsubara frequency $\omega .$ The order parameter $\Delta $
is a sum of $f_{tr}(\omega )$ over positive and negative $\omega $ and
corresponds to a triplet pairing. This triplet component is suppressed by
impurity scattering \cite{Larkin}.

In S/F structures the impurity scattering is rather strong as ferromagnetic
films used in these structures are thin (typically the thickness of the F
films $d$\ is about 20-100 $\bar{A}$) and the elastic scattering at least at
the F surface is strong. Therefore the conventional triplet component would
be strongly suppressed. However there is a special type of the triplet
component which can survive a strong impurity scattering. This triplet
component is described by a condensate wave function $f_{tr}(\omega )$ that
is even in momentum ${\bf p}$ and odd in frequency $\omega .$ This type of
the condensate was first suggested by Berezinskii in 1975 in attempt to
describe the pairing mechanism in superfluid $He^{3}$ \cite{Berezinskii}. It
turned out however that in reality the condensate function in $He^{3}$
is an even function of $\omega $ and an odd function of ${\bf p}$. Later a
possibility to realize the odd (in frequency) triplet superconductivity in
solids was discussed for various models in Refs. \cite{Kirk,Cole,Balat}%
.

The odd in frequency $\omega $ and even in momentum ${\bf p}$ triplet
component in S/F structures differs{\bf \ }from that discussed in the
preceding paragraph. It coexists with the singlet component and the order
parameter $\Delta $ is determined only by the ordinary (BCS) singlet
component $f_{sngl}(\omega )$ even in $\omega $. The triplet component with
nonzero projection of the magnetic moment of Cooper pairs arises as the
result of action of a rotating exchange field on electron spins. This type
of the condensate penetrates the ferromagnet over a long distance of the
order of $\xi _{N}$ (see Eq.(\ref{I1})) provided the period of the
magnetization rotation exceeds $\xi _{N}$.

A S/F structure with an inhomogeneous magnetization has been studied for the
first time in Ref.\cite{BVEprl01}. The authors considered a S/F structure
with a Bloch-type domain wall at the S/F interface in the limit of a short
mean free path ($h\tau <1$). In the domain wall of the thickness $w$ the
magnetization vector was supposed to have the form ${\bf M=}M_{0}\{0,\sin
Qx,\cos Qx\},$ where the $x$-axis is normal to the S/F interface. Outside
the domain wall the magnetization was constant: $M=M_{0}\{0,\sin Qw,\cos
Qw\} $. The condensate function $f_{tr}(\omega ,x)\sim \langle \psi _{\alpha
}(t)\psi _{\alpha }(t^{\prime })\rangle _{\omega }$ was found from the
linearized Usadel equation. It was established that this triplet component
odd in frequency and even in momentum penetrates the magnetic domain wall
over the length

\begin{equation}
{\ }\xi _{Q}=[Q^{2}+2\pi |\omega |/D]^{-1/2}
\end{equation}
It spreads outside the domain wall over distances of the order $\xi _{N}$.
At $x>w$ the vector of the magnetization is fixed so that the first term in
Eq.(\ref{I6}) $Q^{2}$ should be dropped. This triplet component may be
called the long-range triplet component (LRTC). The LRTC may cause a
significant change in the conductance of the Andreev interferometer
consisting of ferromagnetic wires and a superconducting loop. As was shown
in Ref.\cite{BVEprl01}, the conductance variation decreases with increasing
temperature in a monotonic way.

Somewhat later the same problem was considered in the paper \cite{Kadigr}.
In that publication a more complicated situation was discussed, namely, the
case when the width of the domain wall is short compared to the mean free
path. In order to find the condensate function (quasiclassical Green's
function) $f_{tr}(\omega ,x)$ in this case, one needs to solve a more
general Eilenberger equation taking into account a non-homogeneous
magnetization. Unfortunately, the authors of Ref.\cite{Kadigr} did not
manage to solve the Eilenberger equation and therefore restricted themselves
with a rough estimation of the amplitude of $f_{tr}(\omega ,x)$. So, the
problem of calculation of the odd triplet condensate function in the
ballistic regime remained unsolved.

In the present paper we continue studying behavior of the triplet condensate
considering new situations. To be specific, we study the LRTC in S/F
junctions with a spiral magnetic structure in different limits including the
quasi-ballistic one (i.e. $h\tau >1$). This spiral structure may be both an
intrinsic property of a ferromagnet (for example, a helicoidal structure;
see \cite{L-L}) or may just serve as a rough model for magnetic domains.

We consider a system with a conical ferromagnet, that is, we assume that the
magnetization in F rotates in $\{y,z\}$-plane and has a constant component
along the $x$-axis. This type of spin structures is realized, for instance,
in $Ho$ \cite{Ho66} and what we discuss now is a generalization of the
problem considered in \cite{BVEprl01}, where the $x$-component of the
magnetization in F was assumed to be zero. Our study of the proximity effect
in such a S/F system with a spiral magnetic structure is motivated also by
the recent experiment performed on a $Al/Ho$ structure \cite{Sosnin}.

First, solving the Usadel equation, we find the condensate function in the
dirty limit. It turns out that the spatial variation of the LRTC has a
nontrivial dependence on the cone angle $\vartheta $ (see Fig.1). If $%
\vartheta $ is small, the LRTC decays with $x$ exponentially over a distance
of the order of $\xi _{Q}$, but at angles $\vartheta >$ $\sin
^{-1}(1/3)\approx 19^{\circ }$, the LRTC decays in a non-monotonic way. It
oscillates with a period depending on $\vartheta $. These oscillations lead
to oscillations of the critical Josephson current in SFS junction as a
function of the thickness of the F film $2L$ if the thickness $2L$ is
essentially greater than $\xi _{F}$. In this case the Josephson coupling is
only due to the LRTC.

We continue the investigation studying in the third section the LRTC in a
S/F structure with a spiral structure in the limit when the condition $h\tau
>>1$ is valid and the Eilenberger equation must be solved. We analyze
peculiarities of the LRTC in this case. Surprisingly, this case has not been
investigated previously, although it may correspond to a real situation. In
many ferromagnets the exchange energy $h$ is very large so that the product $%
h\tau $ can be arbitrary. In both sections we make an assumption that the
proximity effect is weak, i.e. the amplitude of the condensate function in
the ferromagnet $f$ is small. This assumption is presumably valid in most
cases because there is a strong reflection of electrons at the S/F interface
due to a considerable mismatch in electronic parameters between the
ferromagnet and superconductor \cite{Footnote}. In conclusion we summarize
the obtained results and discuss a possibility of experimental observations
of the predicted behavior of the LRTC.

Note that in Refs. \cite{BVEprl01,Kadigr} and in the present paper a
Bloch-type spiral structure is analyzed. The rotation axis was assumed to be
perpendicular to the S/F interface so that the condensate decays in the
ferromagnet in the direction parallel to this axis (see Fig.1). In this case
the LRTC{\bf \ }arises in the system.

At the same time, one can imagine another, Neel-like type of a spiral
structure with the magnetization vector $M$ that rotates in the plane of the
ferromagnetic film and does not vary in the perpendicular direction. A
solution for this type of the spiral structure has been found in Ref. \cite
{BEL}. The authors considered the case of a thin F film and did not study
the decay length in the direction normal to the film.

The same problem for a thick (infinite) ferromagnetic film was analyzed by
Champel and Eschrig in a recent paper \cite{ChEschrig}. These authors
assumed that the magnetization vector $M$ lies in the plane of the
ferromagnetic film and rotates in this plane being constant in the
perpendicular direction. Surprisingly, they found that there was no LRTC in
this case.

As was shown later \cite{Fominov},\ the absence of LRTC is specific for this
homogeneous spiral magnetic structure. In contrast, the long-range triplet
component arises if the Neel-type spiral structure is non-homogeneous; for
example, if the F film consists of magnetic domains separated by the Neel
walls. In this case the LRTC appears in domain walls and decays in domains
over a long length of the order $\xi _{N}$. Another case where the LRTC
arises is a spin-active S/F interface. Such type of the interface leads to
mixing singlet and triplet component and the triplet component may penetrate
even into a half-metal ferromagnet when the conduction band consists of
electrons with only one direction of spins \cite{Eschrig}.

\section{\protect\bigskip S/F structure with a rotating magnetization. Dirty
limit}

We consider a system shown schematically in Fig. 1. A ferromagnetic film F
is attached to a superconductor S. The magnetization $M$ in the ferromagnet
is assumed to rotate in space and the vector ${\bf M}$ has the form

\begin{figure}
\begin{center}
\includegraphics[width=0.9\textwidth]{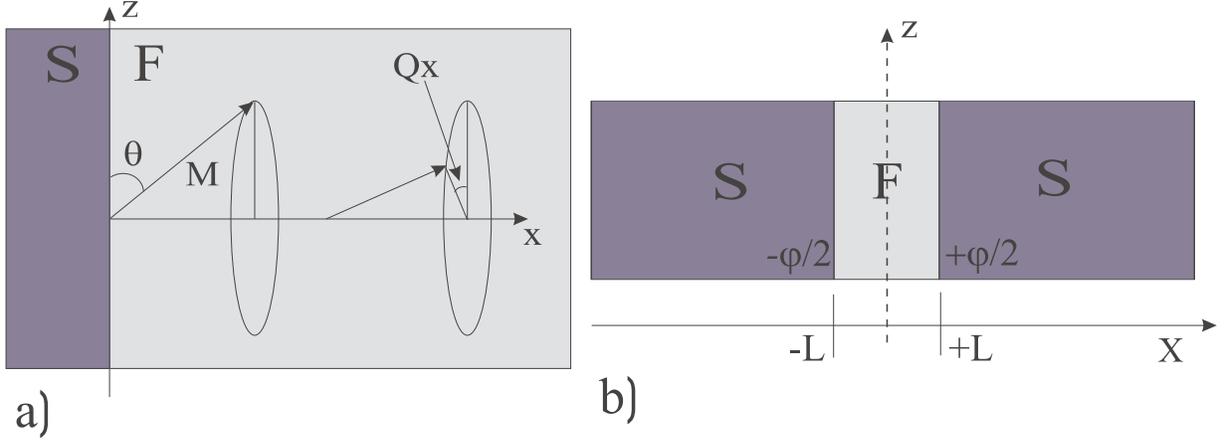}
\end{center}
\large{\caption{Schematic view of the system under consideration. a) S/F structure with a
conical ferromagnet. Magnetization in the ferromegnet F rotates in the
\{y,z\} plane and has a constant projection on the x-axis. b) Josephson
junction with a conical ferromagnet.
}}

\end{figure}

\begin{equation}
{\ }{\bf M=}M_{0}\{\sin \vartheta ,\cos \vartheta \sin Qx,\cos \vartheta
\cos Qx\}  \label{D1}
\end{equation}
This form of the magnetization vector implies that the projection of the $%
{\bf M}$ vector onto the $x$-axis $M_{0}\sin \vartheta $ is constant and the
projection onto the $\{y,z\}$-plane $M_{0}\cos \vartheta $ rotates in space
with the wave vector $Q$. We also assume that the condition

\begin{equation}
{\ }h\tau <<1  \label{D1a}
\end{equation}
is fulfilled, that is, either the exchange energy $h$ is not large or the
collision frequency $\tau ^{-1}$ is high enough. In addition, we assume that
the proximity effect is weak, i.e., \ the condensate function $f$ is small.
The smallness of the $f$ function means the presence of a barrier at the S/F
interface or a big mismatch in electronic parameters of the
superconductor and ferromagnet (the Fermi momenta in F and S differ
greatly). In this case one can linearize the Usadel equation and represent
it in the form \cite{BVEpr01,BVErmp}

\begin{equation}
\partial ^{2}\check{f}/x^{2}-2k_{\omega }^{2}\check{f}+ik_{h}^{2}{\rm sgn}%
\omega \{\cos \vartheta (\left[ \hat{\sigma}_{3},\check{f}\right] _{+}\cos
\alpha (x)+\hat{\tau}_{3}\left[ \hat{\sigma}_{2},\check{f}\right] _{+}\sin
\alpha (x))+\sin \vartheta \hat{\tau}_{3}\left[ \hat{\sigma}_{1},\check{f}%
\right] \}=0,  \label{D2}
\end{equation}
where $\omega =\pi T(2n+1)$, $k_{\omega }^{2}=|\omega |/D$, $k_{h}^{2}=h/D$,
the brackets $[\hat{\sigma}_{3},\check{f}]_{+}$ and $[\hat{\sigma}_{3},%
\check{f}]$ denote anticommutator and commutator. The quasiclassical
condensate Green's function $\check{f}$ is a $4\times 4$ matrix in the
particle-hole and spin space. The Pauli matrices $\hat{\tau}_{i}$ and $\hat{%
\sigma}_{i}$ operate in the particle-hole and spin space, respectively.

The Green function in the bulk of the superconductor is

\begin{equation}
\check{f}_{S}=\hat{\tau}_{2}\hat{\sigma}_{3}f_{S},  \label{D2a}
\end{equation}
with $f_{S}=\Delta /\sqrt{\omega ^{2}+\Delta ^{2}}$. This function describes
an ordinary singlet condensate in conventional superconductors. Assuming
this solution to be valid up to the interface, we use the ``rigid''\
boundary condition at the SF interface ($x=0$) \cite{Zaitsev,Kupriyanov} 
\begin{equation}
\partial \check{f}/\partial x=-\check{f}_{S}/\gamma _{B},  \label{D3}
\end{equation}
where $\gamma _{B}=R_{B}\sigma $, $\sigma $ is the conductivity of the
ferromagnet and $R_{B}$ is the interface resistance per unit area. This
boundary condition is valid provided the ratio $\xi _{h}/\gamma _{B}=\rho
\xi _{h}/R_{B}$ is large, that is, the interface resistance should be larger
than the resistance of the F wire of length $\xi _{h}$.

What we have to do is to solve Eq.(\ref{D2}) in the ferromagnetic region ($%
x>0$) with the boundary conditions (\ref{D3}) at $x=0$. Eq.(\ref{D2}) is a
linear differential equation with coefficients depending on the coordinate $%
x $. This dependence can be excluded if we make an unitary transformation
determined by a matrix $\check{U}$ and introduce a new matrix $\check{f}_{n}$

\begin{equation}
\check{f}=\check{U}\check{f}_{n}\check{U}^{+},  \label{D4}
\end{equation}
where the matrix $\check{U}=\exp (i\hat{\tau}_{3}\hat{\sigma}_{1}Qx/2)$
describes a rotation in the spin and particle-hole space. Substituting\ the
expression (\ref{D4}) into Eq.(\ref{D2}), we get the equation for the new
matrix $\check{f}_{n}$

\begin{equation}
-\partial ^{2}\check{f}/\partial x^{2}+(Q^{2}/2+2k_{\omega }^{2})\check{f}%
+(Q^{2}/2)(\hat{\sigma}_{1}\check{f}\hat{\sigma}_{1})-iQ\hat{\tau}_{3}[\hat{%
\sigma}_{1},\partial \check{f}/\partial x]_{+}-ik_{h}^{2}{\rm sgn}\omega
\{\cos \vartheta \left[ \hat{\sigma}_{3},\check{f}\right] _{+}+\sin
\vartheta \hat{\tau}_{3}\left[ \hat{\sigma}_{1},\check{f}\right] \}=0,
\label{D5}
\end{equation}
For brevity we dropped the subindex $n$. The boundary condition (\ref{D3})
acquires the form

\begin{equation}
\partial \check{f}/\partial x+i(Q/2)\hat{\tau}_{3}\left[ \hat{\sigma}_{1},%
\check{f}\right] _{+}=-\check{f}_{S}/\gamma _{B},  \label{D6}
\end{equation}
As concerns the matrix $\check{f}_{S}$, it does not change (it is invariant
with respect to the transformation (\ref{D4})). In order to solve Eq.(\ref
{D5}) with the boundary condition (\ref{D6}), we represent the matrix $%
\check{f}(x)$ as an expansion in Pauli matrices

\begin{equation}
\check{f}(x)=\sum_{i}\hat{\sigma}_{i}\hat{F_{i}}(x)  \label{D7}
\end{equation}
where $i=0,1,2,3$ and $\hat{\sigma}_{0}$ is the unit matrix. The functions $%
\hat{F_{i}}(x)$ are matrices in the particle-hole space. The spatial
dependence of the functions $\hat{F_{i}}(x)$ is determined by the
exponential functions $\hat{F_{i}}(x)\sim A_{i}\exp (-\kappa x)$, where the
eigenvalues $\kappa $ are determined from the determinant of Eq.(\ref{D5}).
Putting the determinant to zero, we obtain the equation for eigenvalues $%
\kappa $ (see Appendix A)

\begin{equation}
(\kappa ^{2}-Q^{2}-2\kappa _{\omega }^{2})^{2}[(\kappa ^{2}-2\kappa _{\omega
}^{2})^{2}+4\kappa _{h}^{4}\sin ^{2}\vartheta ]+4(Q\kappa )^{2}[(\kappa
^{2}-2\kappa _{\omega }^{2})^{2}+4\kappa _{h}^{4}\sin ^{2}\vartheta
]+4\kappa _{h}^{4}\cos ^{2}\vartheta (\kappa ^{2}-Q^{2}-2\kappa _{\omega
}^{2})(\kappa ^{2}-2\kappa _{\omega }^{2})=0  \label{D8}
\end{equation}

This equation has 4 pairs of roots. However, four of them correspond to
solutions growing in the ferromagnet. As we are interested only in decaying
solutions ($%
\mathop{\rm Re}%
\kappa >0$), we keep four proper roots.

In order to simplify the calculations, we consider a limiting case of a
large exchange field assuming that

\begin{equation}
\kappa _{h}^{2}>>\{Q^{2},\kappa _{\omega }^{2}\}  \label{D9}
\end{equation}
In this limit two eigenvalues of the wave vectors $\kappa $ are large and
equal to

\begin{equation}
\kappa _{\pm }=(1\pm i{\rm sgn}\omega )\kappa _{h}  \label{D10}
\end{equation}
These roots describe a rapid decay of the condensate in the ferromagnet over
an ``exchange length''\ $\kappa _{h}^{-1}$ of the order $\xi _{F}$ (see Eq.(%
\ref{I2})). Only these eigenvalues appear in the case of a homogeneous
magnetization. These large eigenvalues have both real and imaginary parts
and they are equal to each other. This means that corresponding
eigenfunctions decay and oscillate in space on the same scale. The
oscillation of the function $F_{3}$ (singlet component) lead to oscillations
of the critical temperature $T_{c}$ and Josephson current $I_{c}$ in SFS
structures with varying the F film thickness $d$ \cite
{GolubovR,BuzdinR,BVErmp}.

In addition {\bf to} these large values $\kappa _{\pm }$, there are two
other solutions $\kappa _{a,b}$ that describe a long-range penetration of
the triplet component. The singlet component also contains a part decaying
slowly but, as we will see, its amplitude is small in comparison with the
amplitude of the triplet component. These eigenvalues are small: $\kappa
_{a,b}^{2}\approx \max \{Q^{2},\kappa _{\omega }^{2}\}.$ One can find exact
expressions for $\kappa _{a,b}$, but in order to make results more
transparent, we represent solutions in the limit

\begin{equation}
Q^{2}>>\kappa _{\omega }^{2}  \label{D10a}
\end{equation}
In this case Eq.(\ref{D8}) is reduced to the following quadratic equation
for the eigenvalues $\zeta _{a,b}\equiv \kappa _{a,b}^{2}/Q^{2}$

\begin{equation}
\text{\ }\zeta ^{2}-\zeta (1-3\sin ^{2}\vartheta )+\sin ^{2}\vartheta =0
\label{D11}
\end{equation}
with the roots

\begin{equation}
\text{ \ }\zeta _{a,b}=3/2\cdot {\LARGE \lbrack }1/3-\sin ^{2}\vartheta \pm
\cos \vartheta \sqrt{1/9-\sin ^{2}\vartheta }{\LARGE ]}  \label{D12}
\end{equation}

Now, let us consider these eigenvalues as a function of the angle $\vartheta 
$. If the magnetization in the ferromagnet lies in $\{y,z\}$ plane (i.e., $%
\vartheta =0$), we obtain only one root: $\zeta _{a}=1$ (the other root
should be dropped as it corresponds to a solution with the zero amplitude).
This means that the triplet component penetrates the ferromagnet over the
long-range distance $\xi _{LR}\sim Q^{-1}.$ If the value of $Q^{2}$ is
comparable with or less than $\kappa _{\omega }^{2}$, then the
characteristic penetration length of the triplet component is $\xi _{LR}=1/%
\sqrt{Q^{2}+\kappa _{\omega }^{2}}$ (compare with Ref. \cite{BVEprl01}). As
follows from Eq.(\ref{D12}), at small $\vartheta $ the first eigenvalue
equals $\zeta _{a}\simeq 1-4\sin ^{2}\vartheta \approx 1-4\vartheta ^{2}$,
whereas the second root is small and equal to $\zeta _{b}\simeq \sin
^{2}\vartheta \approx \vartheta ^{2}$. Therefore in this case the
exponential decay of the triplet component is slow: $f_{tr}\sim \exp
(-Qx|\vartheta |)$.

Both the eigenfunctions decays exponentially in a monotonic way over a
length much longer than $\xi _{F}$. The situation changes if $\sin \vartheta 
$ exceeds the value $1/3$ ($\theta \gtrsim \sin ^{-1}(1/3)$). In this case
the LRTC oscillates and decays in a non-monotonic way. The period of the
oscillations $\sim 1/Q$ is much longer than $\xi _{F}$. If the magnetization
vector is oriented almost along the $x$-axis ($\sin \vartheta \rightarrow
1,\cos \vartheta \rightarrow 0$), the period of oscillations $\sim Q^{-1}$
is much shorter than the decay length of the LRTC which is equal to $\sqrt{2}%
/(Q\cos \vartheta )>>1/Q$ (see Fig.2).

The amplitudes of the components $\hat{F}(x)_{i}$ are found from the
boundary conditions (\ref{D6}). One can show that the matrices $\hat{F}%
(x)_{i}$ can be represented in the form (see Appendix)

\begin{equation}
\hat{F}(x)_{0,3}=\hat{\tau}_{2}\sum_{k}A_{0,3k}\exp (-\kappa _{k}x)
\label{D13}
\end{equation}

\begin{equation}
\hat{\tau}_{3}\hat{F}(x)_{1,2}=\hat{\tau}_{2}\sum_{k}A_{1,2k}\exp (-\kappa
_{k}x)  \label{D14}
\end{equation}
where the summation is carried out over all eigenvalues: $k=\pm ;a,b.$ In
the approximation, Eq. (\ref{D9}), the amplitudes are

\begin{equation}
\mp A_{0\pm }/\cos \vartheta =\pm iA_{2\pm }/\sin \vartheta =A_{3\pm
}=(f_{S}/2\gamma _{B}\kappa _{\pm })  \label{D15}
\end{equation}

\begin{equation}
-i\tan \vartheta A_{2a,b}=A_{0a,b}=(\zeta _{a,b}-1)/(2i\sqrt{\zeta _{a,b}}%
)A_{1a,b}  \label{D16}
\end{equation}
The amplitude of the long-range triplet component is equal to

\begin{equation}
A_{1a}={\rm sgn}\omega \frac{f_{S}}{\gamma _{B}\kappa _{h}}\frac{\zeta
_{a}\cos \vartheta }{(\zeta _{a}+1)(\sqrt{\zeta _{a}}-\sqrt{\zeta _{b}})}
\label{D17}
\end{equation}
The expression for $A_{1b}$ is obtained from Eq.(\ref{D17}) by permutation $%
a\rightleftarrows b$. These formulas are valid if both $\sin \vartheta $ and 
$\cos \vartheta $ are not too small: $\{\cos \vartheta ,\sin \vartheta
\}>>Q/\kappa _{h}$.

As we mentioned the amplitude $A_{3}$ corresponds to the singlet component
and the amplitude $A_{0}$ describes the triplet component with zero
projection of the magnetic moment on the $z-$axis. In a homogenous case ($M$
is constant in the ferromagnet) this component penetrates the ferromagnet
over a short length of the order $\xi _{F}$. In the case of rotating
magnetization it penetrates the ferromagnetic region over a long distance of
the order of $\min \{1/Q,1/\kappa _{\omega }\}$. The amplitude $A_{2}$
arises only in{\bf \ }the case of a tilted magnetization ($\vartheta \neq 0$%
). One can see from Eqs.(\ref{D15}-\ref{D17}) that in the considered limits
the amplitudes $A_{3\pm },A_{0\pm },A_{2\pm },A_{0a,b},A_{2a,b}$ and $%
A_{1a,b}$ are comparable with each other (other amplitudes are small).
Therefore the LRTC with nonzero projection on the $z-$axis \ is comparable
with the magnitude of the singlet component at the S/F interface $%
(A_{3+}+A_{3-})\sim f_{S}/\gamma \kappa _{h}$. The spatial dependence of the
LRTC amplitude is given by the expression

\begin{equation}
f_{LR}(x)=\frac{f_{S}{\rm sgn}\omega }{\gamma _{B}\kappa _{h}}[\frac{\zeta
_{a}}{\zeta _{a}+1}\exp (-Qx\sqrt{\zeta _{a}})-\frac{\zeta _{b}}{\zeta _{b}+1%
}\exp (-Qx\sqrt{\zeta _{b}})]\frac{\cos \vartheta }{\sqrt{\zeta _{a}}-\sqrt{%
\zeta _{b}}}  \label{D18}
\end{equation}
where $f_{LR}(x)$ is defined in this way: $\hat{F}_{1}(x)=i\hat{\tau}%
_{1}f_{LR}(x)$, $x>>\xi _{F}.$

As it should be, this function is an odd function of $\omega .$ In Fig.2 we
plot the spatial dependence of $%
\mathop{\rm Re}%
(f_{LR}(x)$ for some values of $\sin \vartheta $ and compare it with the
spatial variation of the singlet component.

\begin{equation}
f_{3}(x)=\frac{f_{S}}{2\gamma _{B}}[\frac{1}{\kappa _{+}}\exp (-\kappa
_{+}x)+\frac{1}{\kappa _{-}}\exp (-\kappa _{-}x)]  \label{D19}
\end{equation}

\begin{figure}
\begin{center}
\includegraphics[width=0.9\textwidth]{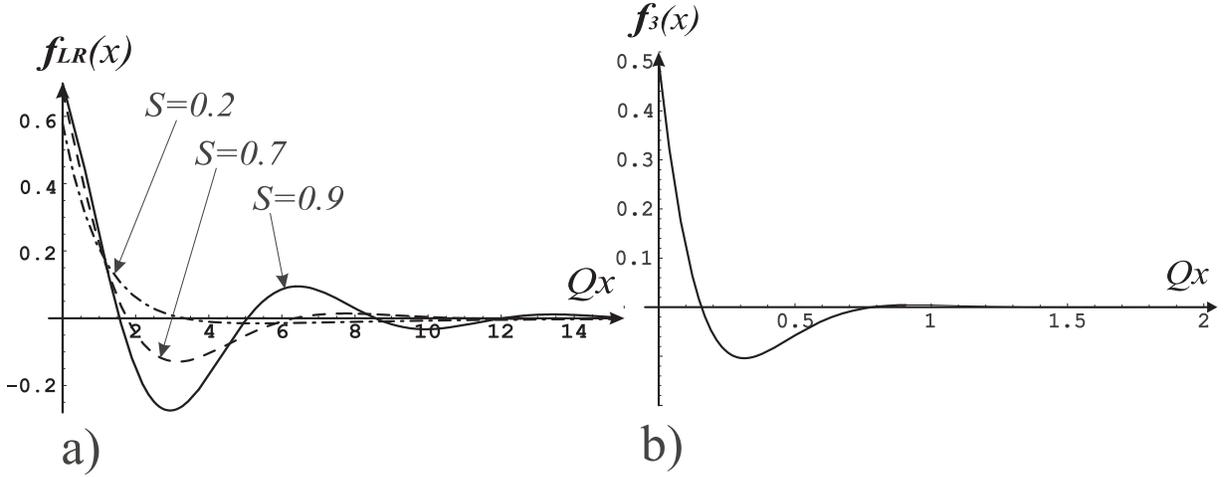}
\end{center}
\large{\caption{Spatial dependence of the real part of the triplet (a) and singlet (b)
component for different angles; $S=\sin \vartheta $ . The singlet component
almost does not depend on $\vartheta $. Both component are normalized to the
quantity ($f_{S}/\gamma \kappa _{h}$ ). The parameter $\kappa _{h}/Q$ is
choosen equal to 5.
}}

\end{figure}

It is seen that the LRTC decays over distances longer than the singlet
component. At small $\sin \vartheta $ the LRTC decays monotonously, but with
increasing $\vartheta $, oscillations of the LRTC arise. The characteristic
scale of the LRTC decay in the considered case (the condition (\ref{D10a})
is fulfilled) is $Q^{-1}$. In the opposite limit of a spiral with a small $Q$
($Q<<\kappa _{\omega }$) the roots $\kappa _{a,b}$ changes: $\kappa
_{a,b}\approx \sqrt{2}\kappa _{\omega }(1\pm iQ\sin \vartheta /\sqrt{2}%
\kappa _{\omega })$. This means that the LRTC decreases exponentially in the
ferromagnet on the length of the order $\xi _{N}$ and oscillates with the
period $(Q\sin \vartheta )^{-1}$. In this case the decay length is shorter
than the oscillation period and therefore there should be no oscillations in
observable quantities. In a general case the characteristic length $l_{tr}$
of the LRTC decay is $l_{tr}\approx \min \{1/Q,\xi _{N}\}.$

In the next Section we discuss a possibility to observe the unusual behavior
of the LRTC and demonstrate that such an observation is realistic.

\section{\protect\bigskip\ Josephson effect in SFS junction}

In this Section we consider a Josephson SFS junction with the same spiral
magnetic structure as in the preceding Section. This junction is shown
schematically in Fig.1b: a ferromagnetic layer with the spiral structure
connects two superconductors where the phases of the order parameter are
equal to $\pm \varphi /2$. We assume again a weak proximity effect and
consider an important case where the Josephson coupling between the
superconductors is only due the LRTC only. This means that the condition

\begin{equation}
\kappa _{h}L>>1  \label{J0}
\end{equation}
should be fulfilled. In this case the singlet component decays fast near the
S/F interfaces and the Josephson coupling is provided by an overlap of the
LRTC.

In order to find the Josephson current $I_{J}$, we need to solve Eq.(\ref{D5}%
) with boundary conditions at $x=\pm L$. The boundary conditions coincide
with Eq. (\ref{D6}), where in this case the matrix $\check{f}_{S}$ at $x=\pm
L$ is replaced by

\begin{equation}
\check{f}_{S}=f_{S}(\hat{\tau}_{2}\cos (\varphi /2)\pm \hat{\tau}_{1}\sin
(\varphi /2))\cdot \hat{\sigma}_{3}  \label{J1}
\end{equation}
The matrix $\check{f}(x)$ is again represented in the form of an expansion
in the Pauli matrices (see Eq.(\ref{D7})), but the ``coefficients''\ $\hat{F}%
(x)_{i}$ in this expansion acquire a somewhat more complicated form. These
matrices (in the particle-hole space) have the structure (see Appendix B)

\begin{equation}
\hat{F}(x)_{0,3}=\hat{\tau}_{2}\sum_{k}A_{0,3k}\cosh (\kappa _{k}x)+\hat{\tau%
}_{1}\sum_{k}{\cal A}_{0,3k}\sinh (\kappa _{k}x)  \label{J2}
\end{equation}
The matrix $\hat{\tau}_{3}\hat{F}(x)_{2}$ has the same structure. As
concerns the matrix $\hat{\tau}_{3}\hat{F}(x)_{1}$, it has a similar
structure but with another spatial dependence

\begin{equation}
\hat{\tau}_{3}\hat{F}(x)_{1}=\hat{\tau}_{2}\sum_{k}A_{1k}\sinh (\kappa
_{k}x)+\hat{\tau}_{1}\sum_{k}{\cal A}_{1k}\cosh (\kappa _{k}x)  \label{J3}
\end{equation}
The coefficients $A_{1k},{\cal A}_{1k}$ can be found from a solution for Eq.(%
\ref{D5}) in a way similar to that in the preceding Section.

The Josephson current (per unit square) is calculated using a general
formula for the condensate current in the dirty case (see, for example, \cite
{BVErmp,Kopnin})

\begin{equation}
I_{J}=i(\pi T/4\rho )\sum_{\omega }Tr\{\hat{\tau}_{3}\cdot \hat{\sigma}_{0}%
\check{f}\partial \check{f}/x\}  \label{J4}
\end{equation}
where $\rho $ is the resistivity of the F metal. Making the transformation (%
\ref{D4}), one can rewrite this formula in terms of the new function $\check{%
f}_{n}$

\begin{equation}
I_{J}=(i\pi T/4\rho )\sum_{\omega }Tr\{(\hat{\tau}_{3}\cdot \hat{\sigma}_{0})%
\check{f}(\partial \check{f}/x+i\hat{\tau}_{3}(Q/2)[\check{f},\hat{\sigma}%
_{1}]_{+})\}  \label{J5}
\end{equation}
We dropped again the index ''$n$''. After simple but somewhat cumbersome
calculations we obtain for $I_{J}$

\begin{equation}
I_{J}=I_{c}\sin \varphi ,\text{ }I_{c}=\pi T(\sigma /2)|Q|\sum_{\omega
,\alpha }\{A_{1\alpha }{\cal A}_{1\alpha }(1-\frac{(\zeta _{\alpha }-1)^{2}}{%
4\sin ^{2}\vartheta })/\sqrt{\zeta _{\alpha }}\}  \label{J6}
\end{equation}
where the summation over $\alpha $ means that $\alpha =a$ and $\alpha =b$.

The amplitudes $A_{1a},{\cal A}_{1a}$ can be found as before. We find (see
Appendix B)

\begin{equation}
A_{1a}=\frac{2\zeta _{a}}{(\zeta _{a}+1)\sinh \theta _{a}}\frac{%
i(A_{3-}-A_{3+})\cos \vartheta }{M}  \label{J7}
\end{equation}

\begin{equation}
{\cal A}_{1a}=\frac{2\zeta _{a}}{(\zeta _{a}+1)\cosh \theta _{a}}\frac{i(%
{\cal A}_{3-}-{\cal A}_{3+})\cos \vartheta }{{\cal M}}  \label{J7a}
\end{equation}
where $M=\sqrt{\zeta _{a}}/\tanh \theta _{a}-\sqrt{\zeta _{b}}/\tanh \theta
_{b},$ ${\cal M}=\sqrt{\zeta _{a}}\tanh \theta _{a}-\sqrt{\zeta _{b}}\tanh
\theta _{b},\theta _{a,b}=\kappa _{a,b}L$. The coefficients $A_{1b},{\cal A}%
_{1b}$ are given by the same formula with replacement $a\rightarrow b$.
Under the condition (\ref{J0}) the coefficients for the singlet component $%
A_{3\pm }$ are described by formulas similar to Eq. (\ref{D15})

\begin{equation}
A_{3\pm }=\frac{f_{S}}{2\gamma _{B}\kappa _{\pm }}\cos (\varphi /2);\text{ \ 
}{\cal A}_{3\pm }=\frac{f_{S}}{2\gamma _{B}\kappa _{\pm }}\sin (\varphi /2)
\label{J8}
\end{equation}

Eqs. (\ref{J6}-\ref{J8}) determine the Josephson critical for the SFS
junction under consideration. In the approximation (\ref{D10a}) we can
perform the summation over $\omega $ and obtain for the critical current

\begin{equation}
\text{ }I_{c}=\frac{2\pi |Q|}{\rho (\gamma _{B}\kappa _{h})^{2}}\Delta \tanh
(\frac{\Delta }{2T})J_{c}  \label{J9}
\end{equation}
Here the coefficient $\gamma _{B}\kappa _{h}$ is assumed to be large. Only
in this case the Usadel equation may be linearized. However the obtained
results are valid qualitatively in the case when this factor is of the order
1. The quantity $J_{c}$ depends only on the angle $\vartheta $ and the
product $QL$. It is equal to

\begin{equation}
\text{ }J_{c}=\frac{\cos ^{2}\vartheta }{M{\cal M}}\sum_{\alpha }\{(\frac{%
\zeta _{\alpha }}{\zeta _{\alpha }+1})^{2}(1-\frac{(\zeta _{\alpha }-1)^{2}}{%
4\sin ^{2}\vartheta })\frac{1}{\sqrt{\zeta _{\alpha }}\sinh 2\theta _{\alpha
}}\}  \label{J10}
\end{equation}

Here the roots $\zeta _{a,b}$ are given by Eq.(\ref{D12}). Since at some
angles the roots $\zeta _{\alpha ,b}$ have an imaginary part, one can expect
that the normalized critical current changes its dependence on $L$ with
varying angle $\vartheta $. First we demonstrate this analytically
considering a limiting case. We assume that the overlap of the condensate
induced by each superconductor is weak. This means that $|\theta _{\alpha
}|>>1$ and therefore $\sinh 2\theta _{\alpha ,b}\approx (1/2)\exp (2\theta
_{\alpha ,b}),$ $\tanh \theta _{\alpha }\approx 1.$ If in addition the angle 
$\vartheta $ is close to $\pi /2,$i.e. $\cos \vartheta <<1$. In this limit
we obtain

\begin{equation}
J_{c}\approx \sqrt{2}\cos \vartheta \exp (-\sqrt{2}QL\cos \vartheta )\sin
(2QL)\text{ }  \label{J11}
\end{equation}
Therefore the normalized critical current $J_{c}(L)$ as a function of $L$
undergoes many oscillations with the period $(\pi /Q)$ on the long decay
length $(\sqrt{2}Q\cos \vartheta )^{-1}$. It turns to zero at $\cos
\vartheta \longrightarrow 0$, but the decay length becomes infinite. We
remind that there is a lower limit on $\cos \vartheta .$ It was assumed that 
$\cos \vartheta $ is larger than the small ratio $Q/\kappa _{h}$. The
maximum of $J_{c}$ is achieved at $\cos \vartheta \approx 1/\sqrt{2}QL$ if $%
QL>>1$.

\begin{figure}
\begin{center}
\includegraphics[width=0.5\textwidth]{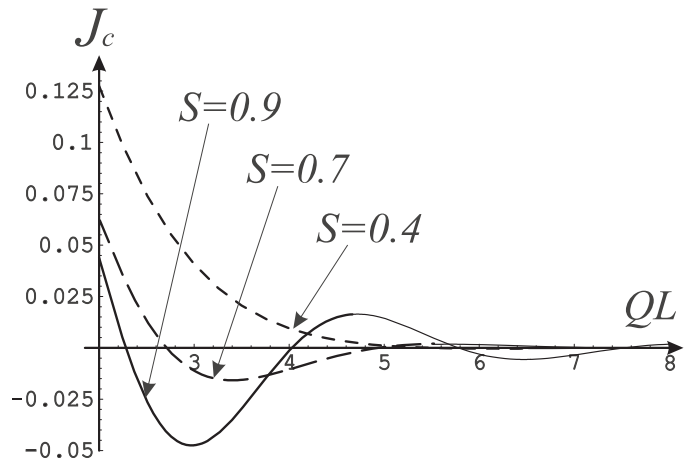}
\end{center}
\large{\caption{ The dependence of the normalized critical Josephson current $J_{c}$ on $QL
$ for various $S=\sin \vartheta $.
}}

\end{figure}

In Fig.3 we plot the dependence of the critical current $I_{c}$ on the
length $L$ for different projection of the magnetization on the $x$-axis. In
accordance with the analysis above, it is seen that at the angle determined
by $\sin \vartheta =1/3$ the decay of $I_{c}$ with increasing $L$ is
accompanied by oscillations. The period of these oscillations is of the
order $Q^{-1}$ and depends on the angle $\vartheta $. These oscillations are
caused by oscillations of the LRTC.

\section{\protect\bigskip\ Strong ferromagnet (quasi-ballistic case)}

In this Section we consider the opposite limit, i.e. the condition

\begin{equation}
h\tau >>1,  \label{C0}
\end{equation}
is assumed to be fulfilled. This case may be realized either in a weak
ferromagnet with a large mean free path or in a strong ferromagnet with a
large exchange energy $h$. For example, this condition is fulfilled for $%
h\approx 1eV$ if the mean free path is longer than $\sim 50$ $\bar{A}$ (we
take the Fermi velocity $v=v_{F\uparrow }\approx v_{F\downarrow }\approx
10^{8}cm/s$). For simplicity we assume that the ${\bf M}$ vector lies in the 
$\{y,z\}$-plane (that is, $\vartheta =0$). In order to find the condensate
function in the ferromagnet, we have to use the more general Eilenberger
equation. We assume again that the proximity effect is weak. The linearized
Eilenberger equation for the condensate matrix function $\check{f}$ reads 
\cite{BVEpr02} (see Appendix A)

\begin{equation}
\mu l\{\hat{\tau}_{3}\partial \check{f}/\partial x+i(Q/2)\left[ \hat{\sigma}%
_{1},\check{f}\right] _{+}\}+2|\omega |\tau \check{f}-ih\tau {\rm sgn}\omega %
\left[ \hat{\sigma}_{3},\check{f}\right] _{+}=\langle \check{f}\rangle -%
\check{f},  \label{C1}
\end{equation}
where $\mu =p_{x}/p,$ $p_{x}$ is the projection of the momentum vector ${\bf %
p}$ on the $x$-axis, $l=\tau v$ is the mean free path, the angle brackets
means the angle averaging. The boundary condition is \cite{Zaitsev}

\begin{equation}
({\rm sgn}\omega )\hat{\tau}_{3}\check{a}=({\rm sgn}\mu )t_{\mu }\check{f}%
_{S}\rfloor _{x=0}  \label{C2}
\end{equation}
where $\check{a}(\mu )=[\check{f}(\mu )-\check{f}(-\mu )]/2$ is the
antisymmetric part of the condensate function$\ $\ and $t_{\mu }=T(\mu )/4,$%
\ $T(\mu )$ is the transmission coefficient of the S/F interface.$\ $In
order to find the solutions of Eq.(\ref{C1}), we represent the matrix $%
\check{f}$ as a sum of antisymmetric and symmetric functions 
\begin{equation}
\check{f}(\mu )=\check{a}(\mu )+\check{s}(\mu )  \label{C3}
\end{equation}
where $\check{s}(\mu )=[\check{f}(\mu )+\check{f}(-\mu )]/2.$ The matrices $%
\check{a}(\mu )$ and $\check{s}(\mu )$ are represented again as a series in
the spin matrices $\hat{\sigma}_{i}$

\begin{equation}
\check{a}=\sum_{i}\hat{a}_{i}\hat{\sigma}_{i},\text{ \ }\check{s}=\sum_{i}%
\hat{s}_{i}\hat{\sigma}_{i}\text{\ }  \label{C4}
\end{equation}
where $i=0,1,3.$ The coefficients $\hat{a}_{i},\hat{s}_{i}$ are, as before,
matrices in the particle-hole space. We substitute these expansions into Eq.(%
\ref{C1}) and single out the symmetric and antisymmetric parts. After some
algebra we obtain for the diagonal matrix elements in the spin space $(%
\check{s})_{11(22)}=\hat{s}_{0}\pm \hat{s}_{3}\equiv \hat{s}_{\pm }$ (see
Appendix C)

\begin{equation}
\hat{s}_{\pm }(x)=\pm \hat{\tau}_{2}t_{\mu }f_{S}\exp (-K_{\pm }x/l\mu )
\label{C6}
\end{equation}
where $K_{\pm }=\alpha _{\omega }\mp i\alpha _{h},$ $\alpha _{\omega
}=1+2|\omega |\tau ,\alpha _{h}=2h\tau {\rm sgn}\omega $. The singlet ($\hat{%
s}_{3}$) component and the triplet ($\hat{s}_{0}$) component with zero
projection of the magnetic moment of Cooper pairs are related to $\hat{s}%
_{\pm }(x):\hat{s}_{3,0}(x)=(\check{s}_{11}(x)\mp \check{s}_{22}(x))/2$. As
is seen from Eq.(\ref{C6}), these components oscillate in space with a short
period of the order of $v/h$ and decay over the mean free path $l$ (if $\tau
T<<1$). When deriving the expression for $\hat{s}_{3,0}(x)$, we assumed that
the condition

\begin{equation}
Ql/|\alpha _{h}|=Qv/h<<1  \label{C7}
\end{equation}
is satisfied, but the relation between the period $Q^{-1}$ of the spiral and
the mean free path may be arbitrary.

Let us turn to the more interesting LRTC $\hat{s}_{1}$. An equation that
describes the LRTC differs considerably from the one for the matrices $%
\hat{s}_{\pm }(x)$. Characteristic wave vectors are much smaller in this
case. For the Fourier transform $\hat{s}_{1}(k)=\int dx\hat{s}_{1}(x)\exp
(ikx)$ this equation has the form

\begin{equation}
\lbrack \alpha _{\omega }^{2}+(Q^{2}(\alpha _{\omega }/\alpha
_{h})^{2}+k^{2})(\mu l)^{2}]\hat{s}_{1}(k)+(l\mu )^{2}(Qk)\hat{\tau}_{3}[%
\hat{s}_{0}+i(\alpha _{\omega }/\alpha _{h})\hat{s}_{3}]=\alpha _{\omega
}\langle \hat{s}_{1}\rangle   \label{C8}
\end{equation}

The matrices $\hat{s}_{0,3}(x)$ are given in the first approximation by Eq.(%
\ref{C6}). Eq.(\ref{C8}) can be solved in a general case, but we are
interested in the behavior of the LRTC at distances much longer than the
mean free path $l$. This means that one has to find the matrix $\hat{s}%
_{1}(k)$ for small $k$: $k<<l$. We find (see Appendix C).

\begin{equation}
\hat{s}_{1}(k)=-\frac{6Q}{k^{2}+K_{Q}^{2}}\frac{\alpha _{\omega }^{2}}{%
\alpha _{h}}\langle \mu ^{2}t_{\mu }\rangle f_{S}i\hat{\tau}_{1}  \label{C9}
\end{equation}
where $\langle \mu ^{2}t_{\mu }\rangle =\int_{0}^{1}d\mu \mu ^{2}t_{\mu }$
and $K_{Q}^{2}=2|\omega |/D+(Q/\alpha _{h})^{2}$. Performing the inverse
Fourier transformation, we find the spatial dependence of the LRTC

\begin{equation}
\hat{s}_{1}(x)=-3\frac{Q}{\alpha _{h}K_{Q}}\langle \mu ^{2}t_{\mu }\rangle
f_{S}\exp (-xK_{Q})i\hat{\tau}_{1},  \label{C10}
\end{equation}
We took into account that $\alpha _{\omega }^{2}\approx 1$. 

This is the main result of this Section. The spatial dependence of the
singlet component $\hat{s}_{3}(x)=(s_{+}-s_{-})/2$ is given by Eq.(\ref{C6})
Comparing Eqs.(\ref{C10}) and (\ref{C6}), we see that the amplitude of the
LRTC is comparable with the amplitude of the singlet component at the S/F
interface. Indeed, if $\sqrt{2\pi T/D}<<Q/\alpha _{h}$, then $K_{Q}\approx
Q/\alpha _{h}$ and the coefficient $(Q/\alpha _{h}K_{Q})$ in Eq.(\ref{C10})
is of the order of $1$.

\begin{figure}
\begin{center}
\includegraphics[width=0.9\textwidth]{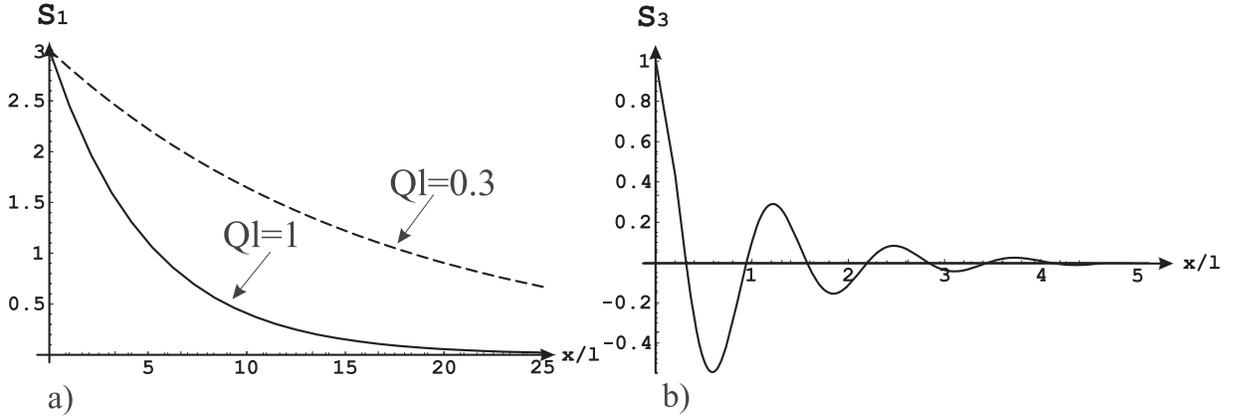}
\end{center}
\large{\caption{Spatial dependence of the real part of the LRTC(a) and singlet (b)
component in quasiballistic case. The singlet component shown for $\mu =1$
does not depend on $Q$. Both component are normalized to the quantity $%
\langle \mu ^{3}t_{\mu }\rangle f_{S}$ and ($t_{\mu }f_{S}$ ), respectively.
The parameter $hv/Q$  is choosen equal to 5.
}}

\end{figure}

In Fig.4 we plot the spatial dependence of the LRTC and the singlet
component. One can see that the singlet component oscillates fast with the
period of the order $v/h=l/h\tau $ (we take the magnitude of $h\tau $ equal
to $5$) and decays over the mean free path $l$. The LRTC decays smoothly
over a length $\sim (Q/h\tau )^{-1}$ (we assumed that $(Q/h\tau )^{2}>2\pi
T/D$).

\section{\protect\bigskip Conclusions}

We considered the odd triplet component of the superconducting condensate in
S/F systems with a spiral magnetic structure. The axis of the spiral is
assumed to be perpendicular to the S/F interface (the Bloch-like spiral
structure). We analyzed both dirty ($h\tau <1$) and clean ($h\tau >1$)
limits. These limits correspond in practice to the cases of weak and strong
ferromagnets.

In the diffusive limit we studied the case of a conical ferromagnet when the
magnetization vector $M$ has the constant projection $M\sin \vartheta $ on
the $x$-axis and rotates around this axis. The condensate amplitude in the
ferromagnet was assumed to be small compared to its amplitude in the
superconductor (a weak proximity effect). In addition, we assumed that the
exchange energy $h$ was larger than such energy scales as $T,DQ^{2}$ (dirty
limit), $Qv$ (clean limit). In this case the penetration length of the
singlet component is much less than that of the LRTC. The singlet component
penetrates the ferromagnet F over a distance of the order $\sqrt{D/h}$ in
the dirty limit and over the mean free path $l$ in the nearly clean limit
(if $\tau T<<1$). In the ballistic case ($\tau T>>1$) the singlet component
decays over distance of order $v/T$. In the nearly clean or ballistic case
the singlet component oscillates fast with the period $v/Q$. The LRTC decays
over a length of the order $\min \{Q^{-1},\xi _{N}\}.$

In conical ferromagnets the LRTC has an interesting nontrivial dependence on
the cone angle $\vartheta .$ At small $\vartheta $ the LRTC decays
exponentially in a monotonic way over the length $1/Q$ (if $Q>>\xi _{N}^{-1}$%
), but at $\vartheta \gtrsim \sin ^{-1}(1/3)\approx 19^{\circ }$ the
exponential decay of the LRTC is accompanied by oscillations. The period of
these oscillations depends on $\vartheta $ so that at $\vartheta
\longrightarrow \pi /2$ the period of oscillations is much smaller than the
decay length. The amplitude of the LRTC is comparable with the amplitude of
the singlet component at the S/F interface. The latter amplitude is
determined by the S/F interface transmittance and decreases with increasing $%
h$.

In the dirty limit we calculated also the critical Josephson current $I_{c}$
for a S/F/S junction with a conical ferromagnet F. It was assumed that the
thickness of the F layer $2L$ is much larger than $\xi _{F}$. Therefore the
Josephson coupling is only due to an overlap of the LRTCs whereas the
overlap of the singlet components induced by superconductors is negligible.
The dependence of $I_{c}$ on the angle $\vartheta $ is determined by the
LRTC: at small $\vartheta $ the critical current $I_{c}$ decreases with $L$
monotonously, but with increasing $\vartheta $ the decay of the function $%
I_{c}(L)$ is accompanied by oscillations. Therefore measurements of the
Josephson critical current in SFS junctions with a conical ferromagnet may
provide useful information about the LRTC.

Note that the triplet component may also exist in magnetic superconductors 
with a spiral magnetic structure (see Ref.\cite{BulR} and references therein). 
However in magnetic superconductors the triplet component coexists 
with the singlet one and, contrary to our case, can not be separated 
from the singlet superconductivity.

We also studied the LRTC in the limit $h\tau >1$ for $\vartheta =0$. In this
case the singlet component decays over a length of the order of the mean
free path $l$ (if $\tau T<<1$). Its amplitude at the S/F interface is
determined by the S/F interface transmittance and does not depend on $h$.
The LRTC penetrates the ferromagnet over a length of the order of $(Q/h\tau
)^{-1}$ (if $Q>(h\tau )/\xi _{N}$). The decay length of the LRTC is longer
than the decay length of the singlet component $l$ provided the condition $%
h>Qv$ is valid.

Note that we neglected the spin-orbit interaction. The latter restricts the
penetration length of the LRTC by the value of the order of $D/8\tau _{s.o.}$
where $\tau _{s.o.}$ is the spin-orbit relaxation time \cite{BVErmp}

We would like to thank SFB 491 for a financial support.

\section{Appendix A. Basic equations; dirty case}

.

\bigskip The Eilenberger equation in a stationary case for the system under
consideration has the form

\begin{equation}
\mu v\frac{\partial \check{g}}{\partial x}-i\epsilon \left[ \hat{\tau}_{3}%
\hat{\sigma}_{0},\check{g}\right] -\left[ \hat{\Delta}\hat{\sigma}_{3},%
\check{g}\right] -ih\{\cos \vartheta (\left[ \hat{\tau}_{3}\hat{\sigma}_{3},%
\check{g}\right] \cos \alpha (x)+\left[ \hat{\sigma}_{2},\check{g}\right]
\sin \alpha (x))+\sin \vartheta \left[ \hat{\sigma}_{1},\check{g}\right] )\}+%
\frac{1}{2\tau }\left[ \langle \check{g}\rangle ,\check{g}\right] =0,
\label{A1}
\end{equation}
where $\mu =p_{x}/p$, $v$ is the Fermi velocity, $\check{g}$ is a $4\times 4$
matrix of the retarded (advanced) Green's functions. The order parameter $%
\hat{\Delta}$ is not zero only in the superconductor and the exchange field $%
h$ differs from zero only in the ferromagnet. If the phase in the
superconductor is chosen to be zero, then $\hat{\Delta}=i\hat{\tau}%
_{2}\Delta $. The angle brackets mean the angle averaging. In the Matsubara
representation the energy $\epsilon $ is replaced by $\omega $: $\epsilon
\rightarrow i\omega $. If the condition $h\tau <<1$ is fulfilled, the
antisymmetric in momentum ${\bf p}$ part of $\check{g}({\bf p})$ is small
and can be expressed through the symmetric part of $\check{g}.$ For the
symmetric part $\check{g}$ one can obtain the Usadel equation (see, for
example, \cite{BVErmp}) that in the Matsubara representation reads

\begin{equation}
D\partial (\check{g}\partial \check{g})/\partial x^{2}+\omega \left[ \hat{%
\tau}_{3}\hat{\sigma}_{0},\check{g}\right] -\left[ \hat{\Delta}\hat{\sigma}%
_{3},\check{g}\right] +ih\{\cos \vartheta (\left[ \hat{\tau}_{3}\hat{\sigma}%
_{3},\check{g}\right] \cos \alpha (x)+\left[ \hat{\sigma}_{2},\check{g}%
\right] \sin \alpha (x))+\sin \vartheta \left[ \hat{\sigma}_{1},\check{g}%
\right] )\}=0,  \label{A2}
\end{equation}
where $D=vl/3$ is the diffusion constant.

In the case of a weak proximity effect one can linearize Eqs.(\ref{A1}) and (%
\ref{A2}). For example, in order to obtain the linearized Usadel equation
for a small condensate function $\check{f}(x)$ in the ferromagnet, we
represent $\check{g}(x)$ in the form

\begin{equation}
\check{g}(x)={\rm sgn}\omega \hat{\tau}_{3}\hat{\sigma}_{0}+\check{f}(x),
\label{A3}
\end{equation}
where the first term is the matrix quasiclassical Green's function of a
normal metal. Linearizing Eq.(\ref{A1}) with respect to $\check{f}(x)$, we
come to Eq.(\ref{D2}).

In order to find solutions for the matrix $\check{f}(x)$, we substitute the
representation of this matrix $\check{f}(x)$ in the form of (\ref{D7}) and (%
\ref{D13}-\ref{D14}) into Eq.(\ref{D5}). As a result we obtain on the
left-hand side of this equation a sum of four terms proportional to the
matrices $\hat{\sigma}_{i}$. Coefficients at each matrix $\hat{\sigma}_{i}$
are matrices in the particle-hole space. The sum of these four terms equals
zero. Therefore we obtain four equations for these coefficients at $\hat{%
\sigma}_{i}$, where $i=0,1,2,3$:

\begin{equation}
(-\kappa ^{2}+Q^{2}+2\kappa _{\omega }^{2})\hat{F}_{0}(\kappa )+2iQ\kappa (%
\hat{\tau}_{3}\hat{F}_{1}(\kappa ))-2i\cos \vartheta \kappa _{h}^{2}\hat{F}%
_{3}(\kappa )=0,  \label{A4}
\end{equation}

\begin{equation}
2iQ\kappa \hat{F}_{0}(\kappa )+(-\kappa ^{2}+Q^{2}+2\kappa _{\omega }^{2})(%
\hat{\tau}_{3}\hat{F}_{1}(\kappa ))=0,  \label{A5}
\end{equation}

\begin{equation}
(-\kappa ^{2}+2\kappa _{\omega }^{2})(\hat{\tau}_{3}\hat{F}_{2}(\kappa
))-2\sin \vartheta \kappa _{h}^{2}\hat{F}_{3}(\kappa )=0,  \label{A6}
\end{equation}

\begin{equation}
-2i\cos \vartheta \kappa _{h}^{2}\hat{F}_{0}(\kappa )+2\sin \vartheta \kappa
_{h}^{2}(\hat{\tau}_{3}\hat{F}_{2}(\kappa ))+(-\kappa ^{2}+2\kappa _{\omega
}^{2})\hat{F}_{3}(\kappa )=0,  \label{A7}
\end{equation}

This system of equations has a nonzero solution if the determinant of the
system is zero. Thus we come to Eq.(\ref{D8}) for four eigenvalues $\kappa
_{k}$. In order to determine the matrices $\hat{F}_{i},$ we have to use the
boundary conditions (\ref{D6}). Since the matrix $\check{f}_{S}$ in the
particle-hole space contains only the matrix $\hat{\tau}_{2}$, the matrices $%
\hat{F}_{0,3}(\kappa )$ and $\hat{\tau}_{3}\hat{F}_{1,2}(\kappa )$ also are
proportional to the matrix $\hat{\tau}_{2}$. Therefore one can write the
expansion (\ref{D13})-(\ref{D14}). The amplitude of each mode are $A_{i,k}$,
where the first index $i$ is related to the spin space and the second $k$
mean the eigenvalues of the wave vectors $\kappa _{k}$ ($k=\pm ,a,b$). From
Eqs.(\ref{A4}-\ref{A7}) one can determine relations between amplitudes $%
A_{i,k}(\kappa _{k})$ for each mode. As follows from Eqs. (\ref{A4}-\ref{A7}%
), the coefficients $A_{i}(\kappa _{k})$ are connected with each other via
Eqs.(\ref{D15})-(\ref{D17}) provided the conditions (\ref{D9}) and (\ref
{D10a}) are fulfilled. The matrices $(\hat{\tau}_{3}\hat{F}_{1}(\kappa _{\pm
}))\approx (2iQ/\kappa _{\pm })\hat{F}_{0}(\kappa _{\pm })$ and $\hat{F}%
_{3}(\kappa _{a,b})\approx -(Q/\kappa _{h})^{2}(z_{a,b}/\sin \vartheta )^{2}%
\hat{F}_{0}(\kappa _{\pm })$ are small compared to other matrices. Using
these relations and substituting the representation (\ref{D7}) into the
boundary condition (\ref{D6}), we obtain four equations for the amplitudes $%
A_{i,k}(\kappa _{k})$ ($i=0,1,2,3$)

\begin{equation}
\sum_{n,\alpha }\{\kappa _{n}A_{0n}+\kappa _{\alpha }A_{0\alpha
}-iQA_{1\alpha }\}=0,  \label{A8}
\end{equation}

\begin{equation}
\sum_{n,\alpha }\{\kappa _{n}A_{1n}+\kappa _{\alpha }A_{1\alpha
}-iQ(A_{0n}+A_{0\alpha })\}=0,  \label{A9}
\end{equation}

\begin{equation}
\sum_{n,\alpha }\{\kappa _{n}A_{2n}+\kappa _{\alpha }A_{2\alpha }\}=0,
\label{A10}
\end{equation}

\begin{equation}
\sum_{n,\alpha }\{\kappa _{n}A_{3n}+\kappa _{\alpha }A_{3\alpha
}\}=(f_{S}/\gamma _{b}),  \label{A11}
\end{equation}
where the summation is performed over the eigenvalues $n=\pm ,\alpha =a,b;$
that is, the indices $n$ and $\alpha $ correspond to the short-range and
long-range eigenfunctions, respectively. Solutions for these equations yield
Eqs.(\ref{D15})-(\ref{D17}).

\section{Appendix B. Josephson effect}

We consider the limit $\kappa _{h}L>>1$ when one can neglect the overlap of
the short-range eigenfunctions corresponding to the eigenvalues $\kappa
_{\pm }$. In this case solutions for $\hat{F}_{0,1}(x)$ may be represented
in the form

\begin{equation}
\hat{F}_{0}(x)=\sum_{n,\alpha }(A_{0n}\hat{\tau}_{2}\pm {\cal A}_{0n}\hat{%
\tau}_{1})\exp (-\kappa _{n}(L\mp x))+A_{0\alpha }\hat{\tau}_{2}\cosh
(\kappa _{\alpha }x)+{\cal A}_{0\alpha }\hat{\tau}_{1}\sinh (\kappa _{\alpha
}x),  \label{B1}
\end{equation}
where $n,\alpha $, as before, are equal to $\pm $ and $a,b.$ The matrix $%
\hat{F}_{2}(x)$ has a similar form. The amplitude of the long-range
component of $\hat{F}_{3}(x)$ is small. This can be seen from Eq.(\ref{A4}).
The matrix $\hat{F}_{1}(x)$ has a different spatial dependence

\begin{equation}
\hat{F}_{1}(x)=\hat{\tau}_{3}\sum_{n,\alpha }(A_{1n}\hat{\tau}_{2}\pm {\cal A%
}_{1n}\hat{\tau}_{1})\exp (-\kappa _{n}(L\mp x))+A_{1\alpha }\hat{\tau}%
_{2}\cosh (\kappa _{\alpha }x)+{\cal A}_{1\alpha }\hat{\tau}_{1}\sinh
(\kappa _{\alpha }x),  \label{B2}
\end{equation}

The first term in Eqs.(\ref{B1}-\ref{B2}) describes the modes fast decaying
from the S/F interfaces at $x=\pm L$ and the second term corresponds to the
LRTC. The coefficients $A_{1n},A_{1\alpha }$ and ${\cal A}_{0n},{\cal A}%
_{0\alpha }$ are connected with each other by Eqs.(\ref{A4}-\ref{A7}) (see
Eqs.(\ref{D15})-(\ref{D16})). The additional terms ${\cal A}_{0,1n,\alpha }%
\hat{\tau}_{1}$ appear because the matrix $\check{f}_{S}$ has changed (see
Eq.(\ref{J1})). In order to find these amplitudes, one has to substitute the
expressions Eqs.(\ref{B1}-\ref{B2}) into the boundary conditions

\begin{equation}
\partial \check{f}/\partial x+i(Q/2)\hat{\tau}_{3}\left[ \hat{\sigma}_{1},%
\check{f}\right] _{+}\rfloor _{x=\pm L}=(f_{S}/\gamma _{b})(\hat{\tau}%
_{2}\cos (\varphi /2)\pm \hat{\tau}_{1}\sin (\varphi /2))\cdot \hat{\sigma}%
_{3},  \label{B3}
\end{equation}

Performing the calculations in this way, we arrive at four equations for the
matrices $\hat{F}_{i}(L)$ (compare with Eqs.(\ref{A8})-(\ref{A11}))

\begin{equation}
\sum_{n,\alpha }\{\kappa _{n}A_{0n}+\kappa _{\alpha }A_{0\alpha }\sinh
\theta _{\alpha }+iQA_{1\alpha }\sinh \theta _{\alpha }\}=0,  \label{B4}
\end{equation}

\begin{equation}
\sum_{n,\alpha }\{(\kappa _{n}A_{1n}+\kappa _{\alpha }A_{1\alpha }\cosh
\theta _{\alpha })+iQ(A_{0n}+A_{0\alpha }\cosh \theta _{\alpha })\}=0,
\label{B5}
\end{equation}

\begin{equation}
\sum_{n,\alpha }\{-i\tan \vartheta \hat{\tau}_{3}\kappa _{n}A_{0n}+i\cot
\vartheta \kappa _{\alpha }A_{0\alpha }\sinh \theta _{\alpha }\}=0,
\label{B6}
\end{equation}

\begin{equation}
\sum_{n,\alpha }\{\kappa _{n}A_{3n}+\kappa _{\alpha }A_{3\alpha }\cosh
\theta _{\alpha }\}=(f_{S}/\gamma _{b})\cos (\varphi /2),  \label{B7}
\end{equation}

We expressed $\hat{F}_{2n,\alpha }$ in terms of $\hat{F}_{0n,\alpha }$
making use Eqs.(\ref{D15})-(\ref{D16}). The corresponding equations for the
coefficients ${\cal A}_{i,n,\alpha }$ may be obtained in a similar way.
These equations coincide with Eqs.(\ref{B4})-(\ref{B7}) if one makes the
replacement $\sinh \theta _{\alpha }\rightleftarrows \cosh \theta _{\alpha }$
and $\cos (\varphi /2)\longrightarrow \sin (\varphi /2).$ In the main
approximation in the parameter $Q/\kappa _{h}$ solutions for these equations
are given by Eqs.(\ref{J7})-(\ref{J8}).

\section{Appendix C. Quasi-ballistic case}

In this Section we represent formulas for the condensate function\ $\check{f}%
(x)$ in the case of a strong ferromagnet or a large mean free path $l$ when
the condition (\ref{C0}) is fulfilled. Substituting Eq.(\ref{A3}) into Eq.(%
\ref{A1}) and performing the transformation (\ref{D4}), we obtain the
linearized Eilenberger equation for the new function $\check{f}_{n}(x)$ in
the ferromagnet (for brevity we drop the subindex $n$)

\begin{equation}
\mu \hat{\tau}_{3}l\partial \check{f}/\partial x+i(Q/2)\mu l\left[ \hat{%
\sigma}_{1},\check{f}\right] _{+}+\alpha _{\omega }\check{f}-i(\alpha _{h}/2)%
\left[ \hat{\sigma}_{3},\check{f}\right] _{+}=\langle \check{f}\rangle ,
\label{Ca1}
\end{equation}
where $\alpha _{\omega }=1+2|\omega |\tau ,\alpha _{h}=2h\tau {\rm sgn}%
\omega .$ In order to solve this equation, we represent the matrix $\check{f}%
(x)$ as a sum of matrices symmetric $\check{s}(x)$ and antisymmetric $\check{%
a}(x)$ in the momentum space

\begin{equation}
\check{f}(x)=\check{s}(x)+\check{a}(x),  \label{Ca2}
\end{equation}

Equations for these matrices can be obtained if we substitute Eq.(\ref{Ca2})
into Eq.(\ref{Ca1}) and split it into the symmetric and antisymmetric in $%
\mu $ parts. We write down, for example, equations for the diagonal elements
of matrices $\check{s}(x)$ and $\check{a}(x)$ in the spin space: $\check{s}%
(x)_{11(22)}=\hat{s}_{\pm },$ $\check{a}(x)_{11(22)}=\hat{a}_{\pm }$. These
equation have the form

\begin{equation}
(K_{\pm }^{2}-\mu ^{2}l^{2}\partial ^{2}/\partial x^{2})\hat{s}_{\pm
}+(Ql\mu )^{2}(K_{\pm }/\alpha _{\omega })\hat{s}_{0}=i(\mu l)^{2}Q[1+K_{\pm
}/\alpha _{\omega }]\hat{\tau}_{3}\partial \hat{s}_{1}/\partial x+K_{\pm
}\langle \hat{s}_{\pm }\rangle ,  \label{Ca3}
\end{equation}

\begin{equation}
K_{\pm }\hat{a}_{\pm }=-(\mu l)[\hat{\tau}_{3}\partial _{x}\hat{s}_{\pm }+iQ%
\hat{s}_{1}],  \label{Ca4}
\end{equation}

The coefficients $K_{\pm }$ are defined in Eq.(\ref{C6}). We solve equations
for $\hat{s}_{\pm }$ assuming that the coefficients $K_{\pm }$ are large,
that is, $\alpha _{h}$ is large. In addition, we assumed that the ratio $%
Ql/\alpha _{h}=Qv/h$ is small. In this case the second term on the left-hand
side and all terms on the right-hand side of Eq.(\ref{Ca3}) can be
neglected. Solving this equation with the boundary conditions (\ref{C2}), in
the main approximation we obtain the expression (\ref{C6}). One can easily
check that the term $\langle \hat{s}_{\pm }\rangle $ is much smaller than $%
K_{\pm }\hat{s}_{\pm }$ provided the quantity $\alpha _{h}$ is large. The
equation for the matrix $\hat{s}_{1}$ can be readily obtained in a similar
way. We get

\begin{equation}
(\alpha _{\omega }^{2}+(\mu l)^{2}[Q^{2}(\frac{\alpha _{\omega }}{\alpha _{h}%
})^{2}-\partial ^{2}/\partial x^{2}])(\hat{\tau}_{3}\hat{s}_{1})-(l\mu
)^{2}iQ\partial _{x}[\hat{s}_{0}(1+(\frac{\alpha _{\omega }}{\alpha _{h}}%
)^{2})+i\frac{\alpha _{\omega }}{\alpha _{h}}\hat{s}_{3}]=\alpha _{\omega
}\langle \hat{\tau}_{3}\hat{s}_{1}\rangle   \label{Ca5}
\end{equation}

The boundary condition for $\hat{\tau}_{3}\hat{s}_{1}(x)$ at $x=0$ requires
that $\hat{a}_{1}=0$ at $x=0$. The expression for the antisymmetric matrix $%
\hat{a}_{1}$ has the form

\begin{equation}
\alpha _{\omega }\hat{a}_{1}=-(l\mu )(\partial (\hat{\tau}_{3}\hat{s}%
_{1})/\partial x+iQ\hat{s}_{0})  \label{Ca6}
\end{equation}
As follows from Eq.(\ref{C6}), at the S/F interface in the main
approximation $\hat{s}_{0}(0)=0$. Therefore the boundary condition for the
matrix $\hat{\tau}_{3}\hat{s}_{1}(x)$ may be written as

\begin{equation}
\partial _{x}\hat{s}_{1}=0.  \label{Ca7}
\end{equation}

Eq.(\ref{Ca5}) can be solved in the following way. In the main approximation
the coordinate dependence of the matrices $\hat{s}_{0,3}(x)$ is given by Eq.(%
\ref{C6}) ($\hat{s}_{0,3}=(\hat{s}_{+}\pm \hat{s}_{-})/2$) and these
functions vary over distances $|l/K_{\pm }|\approx v/h$ that are much
shorter than a characteristic scale for the LRTC variation. Therefore
approximately we can represent a solution for Eq.(\ref{Ca5}) in the form 

\begin{equation}
\hat{s}_{1}(x)=\hat{s}_{1h}(x)+\delta \hat{s}_{1}(x)  \label{Ca8}
\end{equation}
where $\hat{s}_{1h}(x)$ is a short-range part of $\hat{s}_{1}(x)$ which is
determined by: $(\partial /\partial x)\hat{s}_{1h}(x)=iQ\hat{\tau}_{3}[%
\hat{s}_{0}(x)+i(\alpha _{\omega }/\alpha _{h})\hat{s}_{3}(x)]$In particular
the matrix $\delta \hat{s}_{1}(x)$ contains the LRTC. As follows from Eq.(%
\ref{Ca6}) and Eq.(\ref{Ca7}), the boundary condition for the function $%
\delta \hat{s}_{1}(x)$ is 

\begin{equation}
\partial _{x}\delta \hat{s}_{1}(x)=-Q[(\alpha _{\omega }/\alpha _{h})\hat{%
\tau}_{3}\hat{s}_{3}(x)]_{x=0}  \label{Ca9}
\end{equation}
where $\hat{s}_{3}(0)=\hat{\tau}_{2}t_{\mu }f_{S}$. The equation for $\delta 
\hat{s}_{1}(k)$ in the Fourier representation ($\delta \hat{s}_{1}(k)=\int
dx\delta \hat{s}_{1}(x)\exp (ikx)$) can be easily obtained from Eqs.(\ref
{Ca5},\ref{Ca8}-\ref{Ca9}). It has the form 

\begin{equation}
(\alpha _{\omega }^{2}+Q^{2}(\frac{\alpha _{\omega }}{\alpha _{h}}%
)^{2}+(k\mu l)^{2})\delta \hat{s}_{1}(k)=\alpha _{\omega }\langle \delta 
\hat{s}_{1}(k)\rangle +2(l\mu )^{2}Q\frac{\alpha _{\omega }}{\alpha _{h}}%
\hat{\tau}_{3}\hat{s}_{3}(0)  \label{Ca10}
\end{equation}

From this equation one can easily find $\delta \hat{s}_{1}(k)$ 

\begin{equation}
\delta \hat{s}_{1}(k)=\frac{2Ql^{2}}{N(k,\mu )}\frac{\alpha _{\omega }}{%
\alpha _{h}}\hat{\tau}_{3}\{\frac{\alpha _{\omega }}{N_{LR}(k)}\langle \frac{%
\mu ^{2}}{N(k,\mu )}\hat{s}_{3}(0)\rangle +\mu ^{2}\hat{s}_{3}(0)\}
\label{Ca11}
\end{equation}
where $N(k,\mu )=\alpha _{\omega }^{2}+Q^{2}(\alpha _{\omega }/\alpha
_{h})^{2}+(k\mu l)^{2}$ and $N_{LR}(k)=1-\alpha _{\omega }\langle 1/N(k,\mu
)\rangle .$ The behavior of the LRTC $\hat{s}_{1}(x)$ is determined by poles
of the functions $N(k,\mu )$ and $N_{LR}(k).$ The first function has poles
at $k\approx l^{-1}$. These poles determine a variation of the matrix $%
\hat{s}_{1}(x)$ over distances of the order of the mean free path from the
S/F interface. The function $N_{LR}(k)$ has poles at much smaller wave
vectors $k$ which determine a long-range penetration of the triplet
component. Indeed for $k<<l^{-1}$ we have: $%
N_{LR}(k)=(l^{2}/3)[k^{2}+K_{Q}^{2}]$ with $K_{Q}^{2}=2|\omega |\tau
/D+(Q/\alpha _{h})^{2}$. Therefore at these wave vectors $k$ the expression
for $\delta \hat{s}_{1}(k)$ is reduced to Eq.(\ref{C8}).

\end{document}